\documentclass[fleqn,usenatbib]{mnras}

\usepackage{graphicx} 
\usepackage{color}
\usepackage{amssymb}

\def\sss{\scriptscriptstyle}
\def\U{{\sss \!U}}
\def\L{{\sss \!L}}
\def\K{{\sss \!K}}

\def\P{{\sss \!P}}
\def\S{{\sss \!S}}
\def\I{{\sss \!I}}
\def\C{{\sss \!C}}

\def\O{{\sss \!O}}
\def\R{{\sss \!R}}

\def\nur{\nu_\mathrm{r}}

\def\nuL{\nu_\L}
\def\nuU{\nu_\U}
\def\nuK{\nu_\K}

\def\nul{\nuL}

\def\nuP{\nu_\P}
\def\nuISCO{\nu_{\I\S\C\O}}

\title[On formula relating the frequencies of twin-peak QPOs]{{On one-parametric formula relating the frequencies of twin-peak quasi-periodic oscillations} }
\author[T\"{o}r\"{o}k et al.]{Gabriel T\"{o}r\"{o}k$^1$ \thanks{E-mail:
gabriel.torok@gmail.com}, Kate\v{r}ina Goluchov\'{a}$^1$, Eva \v{S}r\'{a}mkov\'{a}$^1$, \newauthor
Ji\v{r}\'{i} Hor\'{a}k$^2$, Pavel Bakala$^1$, Martin Urbanec$^1$ \\
$^1$ Institute of Physics and Research Centre for Computational Physics and Data Processing\\
Faculty of Philosophy \& Science, Silesian University in Opava,  Bezru\v{c}ovo n\'am.~13, CZ-746\,01 Opava, Czech Republic \\
$^2$ Astronomical Institute, Bo\v{c}n\'{i} II 1401/2a, CZ-14131 Praha 4 - Spo\v{r}ilov, Czech Republic
}

\begin{document}

\date{}

\pagerange{\pageref{firstpage}--\pageref{lastpage}} \pubyear{2015}

\maketitle

\label{firstpage}

\begin{abstract}
Twin-peak quasi-periodic oscillations (QPOs) are observed in several low-mass X-ray binary systems containing neutron stars (NSs). Timing analysis of X-ray fluxes of more than dozen of such systems reveals remarkable correlations between the frequencies of two characteristic peaks present in the power density spectra. The individual correlations clearly differ, but they roughly follow a common individual pattern. High values of measured QPO frequencies and strong modulation of the X-ray flux both suggest that the observed correlations are connected to orbital motion in the innermost part of an accretion disc. Several attempts to model these correlations with simple geodesic orbital models or phenomenological relations have failed in the past. {We find and explore a surprisingly simple analytic relation that reproduces individual correlations for a group of several sources through a single parameter. When an additional free parameter is considered within our relation,  it well reproduces the data of a large group of 14 sources. The very existence and form of this simple relation supports the hypothesis of the orbital origin of QPOs and provides the key for further development of QPO models. We discuss a possible physical interpretation of our relation's parameters and their links to concrete QPO models.} 
\end{abstract}


\section{Introduction}
\label{section:intro}

Black holes (BHs) and neutron stars (NSs) represent the accreting compact component in several tens of X-ray binaries. {In low mass X-ray binary systems (LMXBs), the mass transfer from the companion onto the compact object occurs due to the Roche lobe's overflow. 
An accretion disc is formed enhancing these objects through high X-ray luminosity coming from its innermost parts.}  In NS systems, additional strong radiation arises from the disc--NS boundary layer {\citep[e.g.,][]{Lew-etal:book:XrB}}.


The LMXBs exhibit a variability over a large range of frequencies. Their power density spectra (PDS) contain relatively coherent features known as quasi-periodic oscillations {\citep[][and references therein]{kli:2000,bel-etal:2002,mcc-rem:2006,kli:2006}}. Apart from strong, the so-called low frequency (LF) QPOs observed in the range of $0.1-100\,$Hz, there are also the high frequency (HF) QPOs observed in the range of $40-1300\,$Hz. Commonly, for BH and NS sources, HF QPOs attract large attention of theoreticians since their frequencies correspond to orbital timescales in the vicinity of the compact object. The strong indication that the corresponding signal originates in the innermost parts of the disc is also supported by the results of the Fourier-resolved spectroscopy \citep[][]{gilf-etal:2000}. In this context, a large variety of models of the observed fast variability has been proposed {\citep[][and several others]{alp-sha:1985,lam-etal:1985,mil-etal:1998a,psa-etal:1999b,ste-etal:2001,wag-etal:2001,klu-abr:2001,kat:2001,tit-ken:2002,abr-etal:2003c,rez-etal:2003,klu-etal:2004,zha:2004,pet:2005a,bur:2005,cad-etal:2008, muk:2009,stu-etal:2013}}.

The NS LMXBs reveal characteristic pairs of HF QPOs, the so-called twin-peak QPOs. More than dozen of systems exhibit remarkable correlations between their `upper' and `lower' frequencies, $\nuU$ and $\nuL$. In this Letter we focus on these frequencies.  We remind the reader that other properties of each oscillation, such as the rms amplitude $\mathcal{A}$ and quality factor $Q$, strongly vary reaching values of up to $Q\sim250$  and $\mathcal{A}\sim20$ that are much higher than those associated to HF QPOs in BH systems. The variations of the quantities are correlated also with the two QPO frequencies {\citep[e.g.,][]{stra-etal:2002,bar-etal:2005,bar-etal:2005:b,bar-etal:2006,men:2006,kli:2006,tor:2009:AA:,wan-etal:2014}}.

\section{Behaviour and fits of the frequency correlations}
\label{section:correlations}

\begin{figure*}
\begin{center}
a) \hfill {$\phantom{a}$} $\quad\quad\quad\quad\quad\quad$b) \hfill $\quad\quad\quad\quad\quad\quad$c) \hfill {$\phantom{a}$}
\includegraphics[width=.95\linewidth]{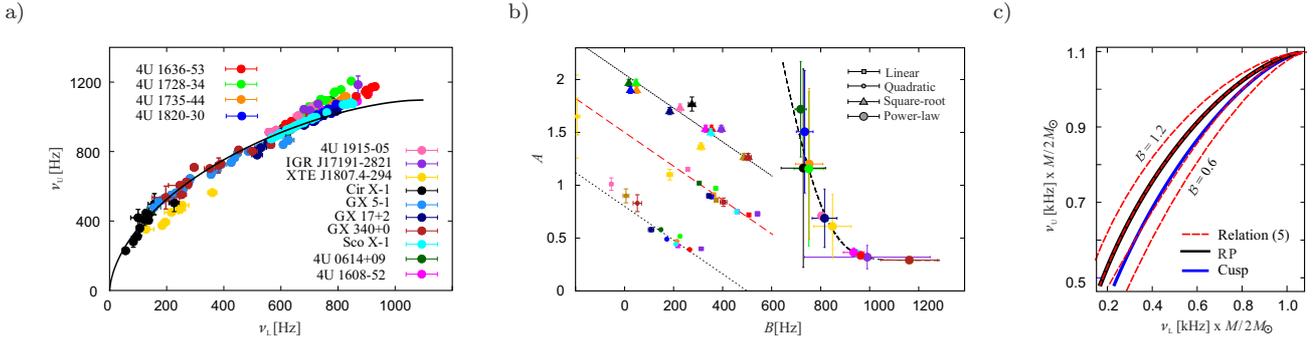}
\end{center}
\caption{The expected and observed correlations between the frequencies of twin-peak QPOs. a) Frequencies of twin-peak QPOs in 14 sources. The black curve indicates the prediction of the RP model assuming a non-rotating NS with $M=2 M_{\sun}$. b) Parameters of various relations fitting the frequencies displayed in panel~a) and the slope intercept anticorrelation found by \citet{abr-etal:2005a, abr-etal:2005b} (red line). For the clarity of drawing, we rescale the parameters of the individual relations as follows. {Linear: $A=a,~B=b$; quadratic: $A=694\,a + 0.06,~B = b - 500 $; square-root: $A=a/30 + 0.2,~B=b + 400$; power-law: $A=42.86\, a + 0.29,~B=500 \, b$. The units of coefficients are chosen such that $A$ is dimensionless and $B$ is given in the units of Hz.} {In two cases, the parameters resulting for Circinus X-1 exceed the displayed range. There is $A = 8.36$ and $B = -298$ for the quadratic relation, and $A = 1.57$ and $B = 273$ for the square-root relation. c) Comparison between the shape of $\nuU(\nuL)$ curves predicted by the RP model (black line), CT model (blue line) and relation  (\ref{equation:the-one}) for $\mathcal{B}\in\{0.6,~0.8,~1.0,~1.2\}$ (dashed red lines). Note that the RP model curve coincides with those given by relation (\ref{equation:the-one}) for $\mathcal{B}=1$ while the CT model curve nearly overlap with those given by relation (\ref{equation:the-one}) for $\mathcal{B}=0.8$.}}
\label{figure:1}
\end{figure*}

The frequency correlations observed in the individual sources clearly differ, but they roughly follow a common individual pattern. This is illustrated in Figure~\ref{figure:1}a which displays twin-peak QPOs observed in 14 different sources. {These include 8 atoll sources, 5 Z-sources, and a milisecond X-ray pulsar. Detailed X-ray timing studies of these objects that have been carried out over the last two decades reveal a large amount of information. A list of individual sources along {with a dozen of related references} is given in Table~\ref{table:1}.} Apart from the datapoints, we include in Figure~\ref{figure:1}a a curve which indicates the trend predicted by the relativistic precession model of HF QPOs - hereafter RP model \citep[see, e.g.,][]{Ste-Vi:1999,ste-etal:2001,bel-etal:2005,tor-etal:2016:ApJ}.

Several  attempts to model the individual observed correlations with simple geodesic orbital models or phenomenological relations have failed in the past \citep[e.g.,][and references therein]{lin-etal:2011,tor-etal:2012,tor-etal:2016:ApJ}. 
In several particular cases, fits are reliable when two free parameters specific for each source are considered {\citep[e.g.,][]{psa-etal:1998,abr-etal:2005a,abr-etal:2005b,zha-etal:MNRAS:2006}}, although there are still numerous clear deviations of data from the expected trend. For instance, fitting by straight lines typically provides reasonable match \citep[but deviation from linear trend is apparent, e.g., in a large amount of data available for the atoll source 4U~1636-53 - ][]{lin-etal:2011,tor-etal:2012}.

It has been noticed by \cite{abr-etal:2005a,abr-etal:2005b} that {the} slope $a$ and intercept $b$ of {the} linear fits obtained for several sources are roughly related as
{\begin{equation}
\label{equation:SIA}
a\approx1.5-0.0015b~\left(\mathrm{assuming}~\nuU=a\nuL+b \right),
\end{equation}}
\noindent
In Figure~\ref{figure:1}b we illustrate the coefficients $a$ and $b$ obtained for each of the sources listed in Table~\ref{table:1}. Within the Figure we furthermore illustrate the coefficients obtained for the other two-parametric fitting relations. {These are namely the quadratic relation, $\nuU=a\nuL^2+b$, the square-root relation, $\nuU=a\sqrt\nuL +b$, and the power-law relation,  $\nuU=b (1\nuL)^{a}$. For each of these relations, including relation (\ref{equation:SIA}), the units of coefficients are chosen such that $a$ is dimensionless while $b$ is given in the units of Hz.} Inspecting Figure~\ref{figure:1}b one can speculate that the frequency correlations within a large group of sources can be described by the means of a single parameter.

{\section{A simple formula reproducing the individual correlations}}

{In the series of works \citep[][]{tor-etal:2010,tor-etal:2012,tor-etal:2016:ApJ} the effective degeneracy between various parameters of several orbital QPO models has been discussed.} 
 Within this degeneracy, each combination of NS mass $M$, angular momentum $j$ and quadrupole moment $q$ corresponds to a certain value of a single generalized parameter $\mathcal{M}$, e.g.,  non-rotating NS mass. It follows that, when these parameters dominate and only non-geodesic effects that do not much vary across different systems are assumed within a given QPO model, one may expect that the correlations can be described by a one-parametric relation,
\begin{equation}
\label{equation:one-parametric}  
\nu_{\L,\,\U}\propto\left(\mathrm{r},\,\mathcal{M}\right) \Rightarrow \nuL=\nuL\left(\nuU,\,\mathcal{M}\right), \, \nuU=\nuU\left(\nuL,\,\mathcal{M}\right),
\end{equation}
{where the common internal parameter r does not appear in the function $\nuU(\nuL)$.} This expectation is in good agreement with the possible degeneracy of the two-parametric frequency relations mentioned in Section~\ref{section:correlations}.

\vspace{0.5cm}

{\subsection{Frequency scaling}}

{Motivated by the above mentioned findings we attempt to model the observed correlations with the following relation,}
\begin{equation}
\label{equation:the-one-most-simple}  
{\nuL= \nuU\left(1 - {\mathcal{B}}\sqrt{1 - \left(\nuU/\nu_0\right)^{2/3}}\right)},
\end{equation}
{where $\nu_0$ represents the highest possible QPO frequency, $\nu_0\geq\nuU\geq\nuL$. In the specific case when it is assumed that $\nu_0$ equals the Keplerian orbital frequency at the innermost stable circular orbit around a non-rotating NS with gravitational mass $M$, it can be expressed in the units of Hz as \citep[e.g.,][]{klu-wag:1985:,klu-etal:1990:}}

\begin{equation}
{\nu_0= \nuISCO=\frac{1}{6^{3/2}}\frac{c^{3}}{2\pi G}\,\frac{1}{{M}}=2198\,\frac{M_{\sun}}{M} = 2198\,\frac{1}{\mathcal{M}}.}
\end{equation}
{Consequently, relation (\ref{equation:the-one-most-simple}) can be written in the form}
\begin{equation}
\label{equation:the-one}  
\nuL= \nuU\left(1 - \mathcal{B}\sqrt{1 - 0.0059\left(\nuU\mathcal{M}\right)^{2/3}}\right).
\end{equation}
{Considering $\mathcal{B}=1$, relation (\ref{equation:the-one}) coincides with the frequency relation implied by the RP model. Moreover, for any constant value} of $\mathcal{B}$, relation (\ref{equation:the-one}) implies $1/\mathcal{M}$ scaling of the QPO frequencies. This means that the $[\nuL,\,\nuU]$ frequency pairs calculated for a certain value of $\mathcal{M}=\mathcal{M}_1$ can be recalculated for another value, $\mathcal{M}=\mathcal{M}_2$, using a simple multiplication,
\begin{equation}
\label{equation:scale}
[\nuL,\,\nuU]_2=[\nuL,\,\nuU]_1\times(\mathcal{M}_1/\mathcal{M}_2).
\end{equation}

{In Figure~\ref{figure:1}c we illustrate the shape of $\nuU(\nuL)$ curves given by relation (\ref{equation:the-one}) and compare them to the predictions of two previously proposed models of QPOs. We assume in this Section for relation (\ref{equation:the-one}) to have either one-parametric or two-parametric form, and in both cases we make a comparison with the data of the individual sources. We then provide a more detailed discussion on the motivation for our choice of the form of relation (\ref{equation:the-one}) in Section~\ref{section:interpretation}.}


\vspace{0.5cm}

{\subsection{The one-parametric relation (pure $1/\mathcal{M}$ scaling)}}
\label{section:one-one}

{Taking into account expectation (\ref{equation:one-parametric}), we attempt to reproduce the data of the 14 sources listed in Table~\ref{table:1} assuming relation (\ref{equation:the-one}) for a fixed value of $\mathcal{B}={0.8}$. The particular choice of $\mathcal{B}={0.8}$ stems from the results of \cite{tor-etal:2012}. At the same time it links relation (\ref{equation:the-one}) to a QPO model introduced by \cite{tor-etal:2016:MNRAS} which deals with marginally overflowing inner accretion tori. See Figure~\ref{figure:1}c for illustration and Section~\ref{section:interpretation} for a more detailed discussion on this matter.}


{The results obtained for the whole set of sources are presented in Table~\ref{table:1} and illustrated in Figure~\ref{figure:2}a. Remarkably, for the sources 1-9 
good agreement is obtained. For each of these sources, there is $0.5<\chi^2/\mathrm{d.o.f.}\lesssim 2$. These sources span (approximately) the range of $\nuL\in(200,~900)$Hz and include the atoll source 4U~1915-05 which itself covers a large range of frequencies, $\nuL\in(200,~800)$Hz. We can also see from Figure~\ref{figure:2}a that the trend observed in each of the sources 10-12 is still matched.} 

The two sources, GX 5-1 and GX 340+0, reveal clear deviations of data from the expected trend (see the last sub-panel of Figure~\ref{figure:2}a). In Table~\ref{table:1} we include relevant $\chi^2$ values along with the values of $\mathcal{M}$ parameter obtained for each of the considered 14 sources. The obtained $\mathcal{M}$ parameter values range from {$\mathcal{M}=0.7$} to {$\mathcal{M}=2.6$} while in most cases there is {$\mathcal{M}\in(1.6,\,1.9)$}.

\vspace{0.1cm}

{\subsection{The consideration of the $\mathcal{B}$ parameter as a free parameter}}

{As a second step, using relation (\ref{equation:the-one}) we attempt to reproduce the data assuming two free parameters ($\mathcal{M}$ and $\mathcal{B}$). The results are again presented in Table~\ref{table:1} and illustrated in Figure~\ref{figure:2}a.}

{Clearly, for each of the 14 sources including GX 5-1 and GX 340+0 (No. 13 and 14), the observed trend is well matched (although there is some scatter of datapoints around the expected curves). In Table~\ref{table:1} we summarize the relevant $\chi^2$ values along with the obtained values of $\mathcal{M}$ and $\mathcal{B}$. The values of $\mathcal{M}$ range from {$\mathcal{M}=1.39$} to {$\mathcal{M}=2.85$} while the values of $\mathcal{B}$ range from {$\mathcal{B}= 0.61$} to {$\mathcal{B}=1.11$} (in most cases there is $\mathcal{B}\in(0.1,0.4)$)}.

\vspace{0.5cm}
\section{Physical interpretation}
\label{section:interpretation}

{We begin this Section by briefly recalling the RP model which is most often discussed in relation to the possibility of hot-spots arising in the innermost accretion region.} This model relates the frequencies of the two observed QPOs to the Keplerian frequency  $\nuU = \nuK$ and the relativistic precession frequency $\nuL = \nuP$ of a slightly perturbed circular geodesic motion that occurs at an arbitrary orbital radius $r$. The precession frequency $\nuP$ equals to a difference between the Keplerian and the radial epicyclic frequency, $\nuP = \nuK - \nur$.


{\cite{tor-etal:2012} have suggested a toy non-geodesic modification of the RP model.  In their paper it is assumed that $\nur$ decreases due to non-geodesic effects by a constant $\mathcal{B}^{*}$ factor.  The expected lower QPO frequency is then given as}
\begin{equation}
\label{equation:RP:modification}
{{\nul}=\nul^* + {\mathcal{B}^{*}}\left(\nuU^*  - \nul^*\right)}
\end{equation}
{where $\nul^*$ and $\nuU^*$ denote the frequencies implied by the non-modified RP model.}
{The authors attempt to model the correlation of QPO frequencies observed in the atoll source 4U~1636-53 and find that there is good agreement between the model and the data for the NS mass $M=1.7M_{\odot}$ and $\mathcal{B}^{*}=0.2$.}

{Assuming relation (\ref{equation:RP:modification}), formulae for the orbital frequencies around non-rotating NS, and $\mathcal{B}^{*}=1-\mathcal{B}$, we obtain relation (\ref{equation:the-one}).}

\begin{figure*}
\begin{center}
\includegraphics[width=0.98\linewidth]{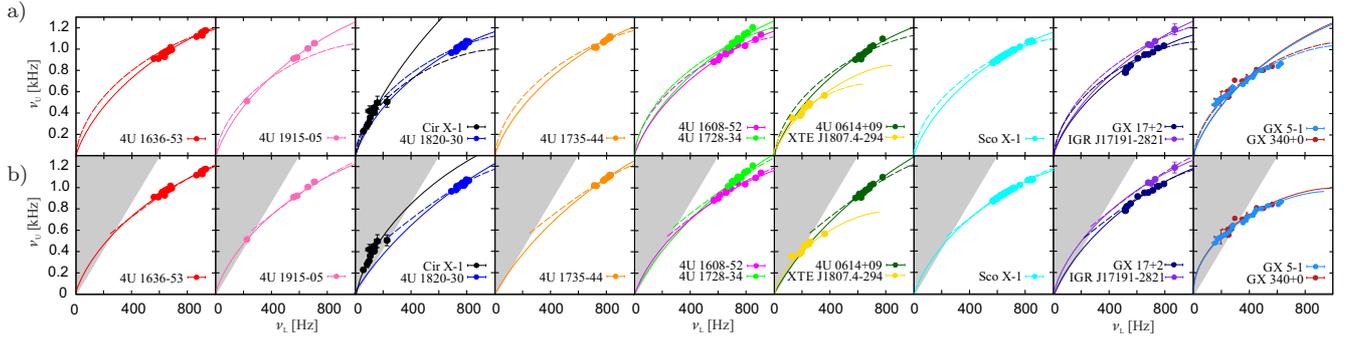}
\end{center}
\flushleft
\vspace{-40ex}
a)
\vspace{15ex}

b) 
\vspace{16.5ex}

\noindent
\caption{{Correlations between the twin-peak QPO frequencies in the individual sources vs. fitting relations. a) Comparison between relation (\ref{equation:the-one}) assuming $\mathcal{B}=0.8$ (solid lines) and the RP model (dashed lines).  b) Comparison between relation (\ref{equation:the-one}) assuming $\mathcal{B}$ parameter as a free parameter (solid lines) and the CT model (dashed lines). The shaded areas denote the range {in which the CT model is not applicable}. For both panels of this Figure, the parameters of the individual fits are summarized in Table~\ref{table:1}.}} 
\label{figure:2}
\end{figure*}

{\subsection{Global modes of fluid motion}}

Oscillations of tori have been studied in the context of QPOs in a large number of works over the past decade {\citep[see, e.g.,][]{rez-etal:2003,abr-etal:2006,sra:2007,Ingram+Done:2010,fragile-blaes:2016,mis-etal:2017,Parth-etal:2017:,ave-etal:2017}}. The recent work of \cite{tor-etal:2016:MNRAS} explores the model of an oscillating torus {(in next CT model). In this model the torus is assumed to fill up the critical equipotential volume forming a cusp.  It is suggested by the model that the twin-peak QPOs are assigned to global modes of cusp-torus fluid motion that may give rise to strong modulation of both the accretion disc radiation and the accretion rate.}{\footnote{{The resulting total observed flux is given by the composition of the emissivity of the boundary layer (radial mode) and the disc (radial and Keplerian mode). Detailed modeling of contributions of particular modulation mechanisms can be important in relation to the recent findings of \cite{Ribeiro-etal:2017} \citep[see also][]{men:2006,gilf-etal:2003}. In this context we note that the considered oscillatory modes can in principle cause also oscillation of the accretion disc corona.}}} {The observed frequency variations are given by the changes of the location $r$ of the torus centre radii very close to the innermost stable circular orbit.  In the paper of \cite{tor-etal:2016:MNRAS} the model predictions are compared to the data of the atoll source 4U~1636-53 obtaining a good match. }

{For a fixed $j$ and given torus location, the frequencies predicted by the model depend solely on the NS mass. Although the related exact dependence $\nuU(\nuL)$ has to be evaluated numerically, we find it can be well approximated by relation (\ref{equation:the-one}) for $\mathcal{B}\approx0.8$. The one-parametric form of relation (\ref{equation:the-one}) investigated in Section \ref{section:one-one} therefore represents the prediction of the CT model (see Figure~\ref{figure:1}c).}

{In Figure~\ref{figure:2}b we directly compare the predictions of the CT model with the data of the individual sources. For sources No. 1-3, 5-7, 10-12, the CT model allows fits comparable to those obtained for relation (\ref{equation:the-one}) and $\mathcal{B}=0.8$. In Table~\ref{table:1} we present  relevant $\chi^2$ values along with the required values of NS mass. In the case of sources No. 4, 8, 9, 13 and 14 (4U 1915-05, Circinus X-1, XTE J1807.4-294, GX 5-1 and GX 340+0) the correlation however cannot be modeled since the observed frequencies $\nuL$ extend below the range of applicability of the approximation of the torus model discussed by \cite{tor-etal:2016:MNRAS} (see the shaded areas in Figure~\ref{figure:2}b).}


{\section{Conclusions}}
\label{section:conclusions}


{The simple formula (\ref{equation:the-one}) well fits the frequencies of twin-peak QPOs within a large group of 14 sources. This match might be of high importance for the twin-peak QPO model identification. The frequency scaling (\ref{equation:scale}), $\nu\propto \mathcal{M}^{-1}$, further supports the hypothesis of the orbital origin of NS HF QPOs since the frequencies of orbital motion scale with the NS mass $M$ as $\nu\propto{M}^{-1}$.  It is of a particular interest that already the one-parametric form of our relation assuming $\mathcal{B}=0.8$ describes several of these sources.  We suggest that this finding represents the NS analogy of the $1/M$ scaling of the 3:2 BH HF QPO frequencies \citep{abr-etal:2004:,mcc-rem:2006,zho-etal:2015}.}

{A detailed physical explanation of the origin of relation (\ref{equation:the-one}) is not yet clear. The recently suggested CT model reproduces the data of 9 sources. For other 5 sources, however, the data cannot be firmly reproduced within the approximation developed so far.} {The results obtained for relation (\ref{equation:the-one}) and the $\mathcal{B}$ free parameter can help improve the model. Larger deviations from the case of $\mathcal{B}=0.8$ can have a direct physical interpretation. The cause of these deviations may be due to a torus thickness different from the cusp value, or, more likely, by further non-geodesic effects acting on the torus formation, such as  NS magnetic field.}

{The above hypothesis agrees with a more general interpretation of relation (\ref{equation:the-one}) in which the $\mathcal{M}$ parameter represents the main parameter reflecting the spacetime geometry given by the NS mass and spin, while the $\mathcal{B}$ parameter reflects the additional stable factors. In this context we note that we have not been able to reproduce the data for any significant group of sources assuming $\mathcal{B}$ as a free and $\mathcal{M}$ as a fixed parameter.} 

\begin{table*}

\caption{{List of sources, references and parameters obtained through data matching. Goodness of fits is formally characterized by $\chi^2$ values \citep[we use the same procedure as][{ displayed errors correspond to  standard errors}]{tor-etal:2012}. The individual columns displaying $\chi^2$ values correspond to relation (\ref{equation:the-one}) {in its one- and two- parametric form}, RP model and CT model. For these two models we assume a non-rotating NS \citep[as discussed in][the spin consideration almost does not improve the fits]{tor-etal:2016:ApJ}. References: {(1)--(3), (10) -- (12)} - \citet{bar-etal:2005, bar-etal:2005:b,bar-etal:2006}, {(4)} - \citet{boi-etal:2000:}, {(5)} - \citet{alt-etal:2010:},  {(6)} -  \citet{Hom-etal:2002:}, {(7)} - \citet{klis-etal:1997:}, {(8)} - \citet{bou-etal:2006:},  {(9)} - \citet{lin-etal:2005:}, {(13)} - \citet{Jon-etal:2000:}, {(14)} - \citet{jon-etal:2002:}. }}
\vspace{-3ex}
 \label{table:1}

\begin{center}
\renewcommand{\arraystretch}{1.0}
{\begin{tabular}{clclllllllll}\hline \hline
Source No./  & Name   & $\mathcal{M}$ &  $\frac{\chi^{2}}{d.o.f.}$ & {$\mathcal{M(B)} $} & {$\mathcal{B}$} & {$\frac{\chi^{2}_{\mathcal{M(B)}}}{d.o.f.}$} & {{$\frac{M_{\mathrm{\R\P}}}{M_{\sun}}$}} & {$\frac{\chi^{2}_{\mathrm{\R\P}}}{d.o.f.}$} &{$\frac{M_{\mathrm{CUSP}}}{M_{\sun}}$} &{$\frac{\chi^{2}_{\mathrm{CUSP}}}{d.o.f.}$}  & Data- \\
Type$^{a}$  &&& &&& &  & &  && points \\  \hline
1/A & 4U 1608-52  & {1.80$^{\pm 0.01}$} &  {1.6} & {1.79$^{\pm  0.04}$} & {0.79$^{\pm  0.03}$} & {1.7}&{1.94}& {10.1} & {1.74$^{\pm 0.01}$} & {1.9}  &12\\ 
2/A & 4U 1636-53  & {1.70$^{\pm  0.01}$}&  {2.0} & {1.70$^{\pm  0.01}$} & {0.8$^{\pm  0.01}$} &{2.1}&{1.79}& {17.4} & {1.69$^{\pm 0.01}$} & {3.4} &22\\ 
3/A & 4U 1735-44  & {1.69$^{\pm 0.01}$} &  {2.1} & {1.48$^{\pm  0.10}$}&{0.61$^{\pm  0.06}$}&{1.0} &{1.81} & {5.1} & {1.66$^{\pm 0.01}$}&  {1.4} &8\\ 
4/A & 4U 1915-05  & {1.58$^{\pm 0.03}$} &  {0.8} & {1.65$^{\pm  0.03}$}&{0.82$^{\pm  0.01}$}&{0.2} &{2.09}& {28.6} &{$-^b$}& {$-^b$} &5\\ 
5/A & IGR J17191-2821  & {1.58$^{\pm 0.02}$} &  {0.6} & {1.63$^{\pm  0.20}$}&{0.85$^{\pm  0.2}$}&{0.8} &{1.76}& {0.6} & {1.52$^{\pm 0.02 }$}& {0.6} & 4\\ 
6/Z & GX 17+2  & {1.89$^{\pm 0.02}$} &  {1.2} &{1.77$^{\pm  0.07}$}&{0.72$^{\pm  0.04}$}&{ 0.8} &{2.08} &  {5.5} &{1.83$^{\pm 0.02 }$}& {0.9} & 10\\ 
7/Z & Sco X-1  & {1.82$^{\pm 0.01}$}  & {1.0} &{1.81$^{\pm  0.01}$}&{0.8$^{\pm  0.01}$}&{1.0} &{2.0} & {24.2} &{1.76$^{\pm 0.01 }$}& {2.3} & 39\\ 
8/Z & Cir X-1  & {0.74$^{\pm 0.10}$} &  {1.2} &{1.42$^{\pm  0.5}$}&{0.89$^{\pm  0.06}$}&{1.1} &{2.23}& {1.3} &{$-^b$}& $-^b$ & 11\\ 
9/P & XTE J1807.4-294  & {2.61$^{\pm 0.11}$} &  {0.8} &{2.85$^{\pm  0.25}$} &{0.86$^{\pm  0.07}$}&{0.8}&{3.27}& {1.4} &{$-^b$}& $-^{b}$ &7\\
10/A & 4U  1728-34  & {1.57$^{\pm 0.01}$} &  {3.2} & {1.35$^{\pm  0.12}$}&{0.65$^{\pm  0.06}$}&{2.5} &{1.74}& {5.7} &{1.51$^{\pm 0.01 }$}&  {2.8} &15\\ 
11/A & 4U 0614+09  & {1.71$^{\pm  0.02}$} &  {5.1} & {1.39$^{\pm  0.06}$} & {0.62$^{\pm  0.02}$}& {1.1} &{1.90} & {14.7} & {1.65$^{\pm 0.01}$} & {3.4} &13\\ 
12/A & 4U 1820-30  & {1.81$^{\pm 0.01}$} &  {9.3} & {1.53$^{\pm  0.07}$}&{0.58$^{\pm  0.03}$}&{3.2} &{1.93}& {24.2} & {1.78$^{\pm 0.01}$}& {6.4} &23\\ \hline
13/Z & GX 340+0  &  {1.62$^{\pm {0.08}}$}  & {4.2} &{2.23$^{\pm  0.10}$}&{1.10$^{\pm  0.08}$}&{1.6}&{2.07}& {1.8} &{$-^b$}& $-^b$ & 12\\ 
14/Z & GX 5-1  & {1.65$^{\pm {0.10}}$}  & {16.7} &{2.31$^{\pm  0.04}$}&{1.11$^{\pm  0.02}$}&{1.5} &{2.13} & {3.1} & {$-^b$}&  $-^b$ & 21\\ 
\hline
\multicolumn{12}{l}{\parbox[t]{17cm}{$^a$ A\,-\,atoll, Z\,-\,Z, P\,-\,pulsar. 
 $^b$ The observed frequencies extend below the expected range of physical applicability of CT model discussed by \citet{tor-etal:2016:MNRAS}.
}}
\vspace{-3ex}

\end{tabular}}

\end{center}
\end{table*}

\section*{Acknowledgments}

{We would like to acknowledge the Czech Science Foundation grant No. 17-16287S,  the INTER-EXCELLENCE project No. LTI17018 supporting collaboration between the Silesian University in Opava (SU) and Astronomical Institute in Prague (ASU), and the internal SU grant No. SGS/15/2016. We thank to the anonymous referee for his/her comments and suggestions that have significantly helped to improve the paper. We are grateful to Marek Abramowicz (SU) and Omer Blaes (University of California in Santa Barbara - UCSB) for useful discussions. Last but not least, we would like to acknowledge the hospitality of UCSB, and to express our thanks to concierges of Ml\'{y}nsk\'{a} hotel in Uhersk\'{e} Hradi\v{s}t\v{e}, Czech Republic for their participation in organizing frequent workshops of SU and ASU.}

\label{lastpage}

\end{document}